\documentclass[letterpaper, 10 pt, conference]{ieeeconf}  

\IEEEoverridecommandlockouts                              

\usepackage{caption,subcaption}
\usepackage{comment} 
\usepackage{graphicx}
\usepackage{siunitx}
\usepackage{amsmath} 
\usepackage{amssymb}  
\usepackage{color,xcolor}
\usepackage{cite}
\usepackage{enumerate}
\DeclareMathOperator{\e}{\mathrm{e}}

\newtheorem{proposition}{Proposition}
\newtheorem{remark}{Remark}
\newtheorem{lemma}{Lemma}
\newtheorem{theorem}{Theorem}
\newtheorem{corollary}{Corollary}

\title{\bf\LARGE Pulsatile Endocrine Regulation with a Releasable Hormone Pool: Impulsive Feedback Model}

\author{ Anton V. Proskurnikov and  Alexander Medvedev
\thanks{*AM was partially supported by the Swedish Research Council under
grant 2024-04943.}
\thanks{Alexander Medvedev [{\tt\small alexander.medvedev@it.uu.se}] is with Department of Information Technology, 
        Uppsala University, SE-752 37 Uppsala, Sweden.}%
\thanks{Anton V. Proskurnikov  [{\tt\small anton.p.1982@ieee.org}] is with the Department of Electronics and Telecommunications, Politecnico di Torino, Turin, Italy, 10129.}%
}

\begin{document}

\maketitle

\begin{abstract}
An impulsive model of endocrine regulation equipped with frequency- and amplitude-modulated pulsatile feedback is extended to study the dynamical effects  pertaining to limitations due to the releasable hormone pool. The model describes a three-hormone regulatory loop and is applicable to endocrine systems such as the reproductive, stress, and growth hormone axes. 
Conditions guaranteeing the existence and uniqueness of a periodic solution with a single feedback-triggered impulse per period (a 1-cycle) are derived, together with monotonicity properties of its characteristics. An explicit Jacobian formula is obtained to assess local orbital stability. Model analysis demonstrates that a reduced hormone-pool replenishment rate produces a lower-amplitude, higher-frequency 1-cycle. This behavior is consistent with experimentally observed alterations in testosterone regulation in aging human males.
\end{abstract}
\section{Introduction}
Restricting the control signal to be a sequence of impulses (Dirac deltas) gives rise to impulsive control~\cite{GC98,HCN14}. The weights and timing of the impulses are manipulated by the controller to achieve the desired plant behavior. In contrast with conventional sampled-data control, that makes use of high sampling rates to mimic continuous (analog) control, impulsive control accomplishes the control objective by exerting sparse intermittent  action.

Impulsive control is a common approach in engineering and widely used  for dosing of additives in process industry, spacecraft orbit correction, and  for implementing gaits in legged robots. It is also ubiquitous in natural biological systems, for instance in pulsatile endocrine regulation \cite{WTT10}.

Due to the autonomy of an organism, all the hormones participating in an endocrine loop (or axis) have to be produced in necessary quantities (autopoiesis). Hormone deficiency causes a systemic disruption to the organism's homeostatic balance via 
e.g. failure of target cell activation and breakdown of feedback loops \cite{VKL09}.

The regulation of testosterone (Te)\footnote{We use the abbreviation Te instead of the commonly accepted T to avoid confusion with the period of a periodic solution. } production by Gonadotropin-Releasing Hormone (GnRH) is mediated through the Hypothalamic-Pituitary-Gonadal (HPG) axis.  Specialized neurosecretory neurons within the hypothalamus synthesize and release GnRH in pulsatile bursts, typically occurring every 60 to 90 minutes in the human male. Each GnRH pulse stimulates the anterior pituitary gland to release a corresponding pulse of luteinizing hormone (LH) into the systemic circulation. LH travels through the bloodstream to the testes, where it binds to LH receptors, stimulating testosterone production. Rising Te levels  exert negative feedback on  the hypothalamus, suppressing the secretion of GnRH and thereby reducing further Te production \cite{VKL09}.


The releasable pool of GnRH  acts as the primary gating mechanism for the entire HPG axis, directly controlling the timing and amplitude of sex hormone secretion. The readily releasable pool consists of GnRH vesicles docked at the nerve terminals in the hypothalamus. The quantity of GnRH available in this immediate pool determines the amplitude of the GnRH pulse sent to the pituitary gland. A larger pulse triggers a larger release of LH, which subsequently drives up the production of sex hormones in the gonads.

The mobilization of GnRH into the releasable pool is highly sensitive to  starvation, extreme exercise, or high stress. Cortisol or low leptin inhibit the transport of GnRH vesicles to the releasable pool, effectively shutting down reproductive hormone production when the body cannot support  optimal health.

Aging fundamentally disrupts the pulsatile secretion of GnRH, shifting it from a highly synchronized, rhythmic cycle to a more irregular and damped output. The shift toward a lower amplitude and higher frequency of GnRH and downstream LH pulses is a classic feature of male reproductive aging. Research shows a 30\% to 50\% increase in GnRH pulse frequency in older men compared to young men \cite{VKL09}.  The size of each LH burst decreases significantly, leading to   pulses of lower amplitude.

Interestingly, in middle-aged male mice, reprogramming the endogenous GnRH pulses via an optogenetic approach revealed that increasing GnRH pulses frequency causes LH excess and aging acceleration, while lowering the frequency of and stabilizing GnRH pulses can slow down aging~\cite{WWK2021}. 

Several endocrine regulation models explicitly incorporate a releasable pool  of GnRH or its immediate downstream targets to describe secretion dynamics \cite{VPL22}. Some of them model the hypothalamic GnRH neurons themselves, and other portray the pituitary response driven by GnRH pools.

A model addressing the releasable pool of GnRH along with its replenishment and depletion is  proposed in~\cite{Lehnert2019}.
It describes the electrical and hormonal processes within a single hypothalamic GnRH neuron and limits its scope to only autocrine feedback mechanism regardless of Te regulation.  A pulse-modulated feedback model that describes how the amplitude and frequency of GnRH pulses are modulated by the Te serum  concentration in the human male is introduced in \cite{MCS06,Aut09}. It possesses hybrid dynamics, allows analytical treatment, and explains clinically observed hormone profiles well \cite{PMAM14,RM22}. The model is also known as the Impulsive Goodwin's Oscillator (IGO) and is generalized to an arbitrary number of the interacting  substances in \cite{PRM24}.

In contrast to \cite{PRM24}, the present work augments the original IGO with an additional state variable representing the amount of GnRH stored in a releasable hormone pool. This extension destroys the cascade structure of the continuous-time dynamics of the IGO and therefore falls outside the mathematical framework considered in \cite{PRM24}. Moreover, the amplitude modulation function now depends not only on the output of the endocrine axis, namely the Te concentration, but also on the amount of releasable GnRH. A related setting, in which both the amplitude and frequency of hormone pulses depend on the full state vector, was investigated in \cite{Churilov2020}; however, the focus there was on the stability of $m$-cycles.


The contributions of the paper stand as follows. First, a mathematical model describing the mechanism of pulsatile endocrine regulation and accounting for the finite capacity of a releasable hormone pool is proposed. Secondly, conditions for existence and uniqueness of 1-cycle are obtained. Thirdly, for the model at hand, it is proven that a reduction in release hormone production leads to weaker but more frequent release events. This is a clinically observed phenomenon in the Te regulation in the aging human male.

\section{Model equations}

We augment the impulsive Goodwin oscillator portraying Te regulation in the male with a finite release
hormone pool of GnRH. The original formulation in \cite{MCS06,Aut09} assumes infinite availability of the release hormone. The additional state $x_r$ represents the hormone amount instantly
available for secretion. Between consecutive release events, the pool is
continuously replenished by hormone synthesis and depleted by degradation,
whereas at each release event a fraction of the available hormone is
transferred to the downstream endocrine cascade.

\paragraph*{Releasable hormone pool}
Between two consecutive pulses, the available GnRH amount $x_r\ge 0$
is described by 
\begin{align}\label{eq:pool}
    \dot x_r(t)&= -a_r x_r(t)+v,
\end{align}
where  $v>0$ is the rate at which newly synthesized GnRH enters the
releasable pool, and $a_r>0$ is its degradation rate. In the absence of
release events, the pool approaches the equilibrium value $v/a_r$.
The corresponding degradation time constant, and half-life are
$1/a_r$ and $\ln(2)/a_r$, respectively. 

Notice that, unlike the states of the downstream model, which represent hormone
concentrations, $x_r$ represents the amount of releasable hormone. This
choice is rational because the quantity removed from the pool at a
release event has the direct interpretation of a hormone dose secreted by the hypothalamus into systemic circulation.

\paragraph*{Downstream endocrine dynamics}

Between release events, and following the classical continuous Goodwin/Smith model \cite{Smi80,Smi83}, the effect of secreted GnRH is described by the
linear  cascade of three compartments
\begin{equation}\label{eq:model_c}
\begin{aligned}
   \dot x_1(t)&=-a_1x_1(t), \\
   \dot x_2(t)&=-a_2x_2(t) + g_1 x_1(t),\\
   \dot x_3(t)&= -a_3x_3(t)+ g_2 x_2(t),\\
   y(t)&=x_3(t),
\end{aligned}
\end{equation}
where $a_i>0$ are degradation or clearance rates and $g_1,g_2>0$ are
transfer coefficients. Traditionally, $x_i$, $i=1,2,3$, are the serum concentrations of GnRH, LH, and T, respectively.
The output $y$ is the feedback variable
regulating both the amount of hormone released and the timing of the
next pulse.

\paragraph{Pulse timing}
Let $t_k$ denote the $k$-th GnRH release time. The time interval until the next
release is determined by the downstream hormone level according to
\begin{align}\label{eq:timing}
    t_{k+1}=t_k+T_k,\quad T_k\triangleq\Phi(y(t_k)),
\end{align}
where\footnote{Henceforth, $\mathbb{R}_+=[0,\infty)$.} $\Phi:\mathbb{R}_+\to\mathbb{R}_+$ is the frequency-modulation
function. Consistently with negative endocrine feedback, 
$\Phi$ is assumed non-decreasing: an increased downstream hormone concentration
postpones the subsequent release event and allows the output $y$ to subside due to hormone degradation.

\paragraph{Hormone release}

At the time instant $t_k$, an amount 
\begin{equation}\label{eq:lambda}
\lambda_k\triangleq F\bigl(x_r(t_k^-),y(t_k^-)\bigr),
\end{equation}
is released from the hormone pool and reaches the first compartment of the downstream cascade: 
\begin{equation}
\begin{aligned}\label{eq:jump}
    x_r(t_k^+)&=x_r(t_k^-)-\lambda_k,\\
    x_1(t_k^+)&= x_1(t_k^-)+\beta \lambda_k\\
    x_2(t_k^+)&=x_2(t_k^-),\\
    x_3(t_k^+)&=x_3(t_k^-).
\end{aligned}
\end{equation}
Here $F:\mathbb{R}_+\times \mathbb{R}_+\to\mathbb{R}_+$ is the amplitude modulation function, and
$\beta>0$ is a conversion coefficient between the amount of hormone
released from the pool and the resulting concentration increment. Hence, if $x_r$ is measured as an amount of
hormone and $x_1$ as a concentration, then $\beta$ has the dimension of
an inverse volume.

A natural admissibility requirement, ensuring that the amount of released hormone does not exceed the available amount of hormone in the pool, is
\begin{gather}
    0<F(\xi,\theta)\leq \xi, \qquad \forall\,\xi>0,\theta\geq0,\label{eq:F-admissible}\\
    F(0,\theta)=0,\qquad\forall\theta\geq 0,\notag
\end{gather}
which guarantees that the released dose $\lambda_k$ is strictly positive except for the degenerate case where $x_r(t_k^-)=0$. Notice
that closed-loop system has no equilibria, since at an equilibrium one would have $x_r=v/a_r>0$, which corresponds to impulses of strictly positive amplitude. For the forthcoming uniqueness analysis, stronger conditions are imposed
\begin{equation}\label{eq:F-monotone}
    F(0,\theta)=0,\quad
    0<F'_\xi(\xi,\theta)\leq1,\quad
    F'_\theta(\xi,\theta)\leq0.
\end{equation}
The first two conditions imply~\eqref{eq:F-admissible}, while
$F'_\theta\leq0$ represents negative feedback with respect to the amplitude: increasing the downstream
hormone level cannot increase the amount released at the next pulse. 


A simple example of a modulation function satisfying these requirements is
\begin{equation}\label{eq:F-product}
    F(\xi,\theta)=\xi\,\bar F(\theta),
    \qquad
    0\leq\bar F(\theta)\leq1,
\end{equation}
with $\bar F$ nonincreasing and strictly positive. Then, $\bar F(\theta)$ is interpreted as the fraction of the currently available hormone pool to be released in a pulse. We do not, however, restrict the subsequent
analysis to this particular multiplicative form.

\section{1-cycles: Existence and Uniqueness}

Consider periodic rhythms in \eqref{eq:model_c}, caused by the impulsive feedback. The simplest periodic solution is characterized by a single  GnRH release event per (least) period; such periodic solutions are referred to as 1-cycles~\cite{Aut09,MPZh18,ZCM12b}.

Denote the state of  \eqref{eq:model_c}  and the  state of the augmented system (including the pool state) by
\[
X=\begin{bmatrix}
    x_1 &x_2 &x_3
\end{bmatrix}^\top,\quad \bar X=\begin{bmatrix} x_r &x_1 &x_2 &x_3 \end{bmatrix}^{\top},
\]
respectively. Let
\[
    X_k=X(t_k^-),\quad\bar X_k=\bar X(t_k^-),\quad y_k=y(t_k^-),
\]
be the corresponding states immediately prior to the $k$-th hormone
release event. 
The  release hormone amount and the inter-pulse interval are found as follows:
\begin{equation}\label{eq:lambda-T-k}
    \lambda_k
    =F(x_{r,k},y_k),
    \quad
    T_k
    =\Phi(y_k).
\end{equation}

The 1-cycle can be equivalently  defined as a solution that reproduces the full state after every release event, that is,
the sequences $\bar X_k,\lambda_k,T_k$ are constant:
\[
\begin{gathered}
    \bar X_{k+1}=\bar X_k=\bar X^*=\begin{bmatrix}
        x_r^* &x_1^* &x_2^* &x_3^*
    \end{bmatrix}^{\top}
    \quad\forall k=0,1,\ldots,\\
    \lambda_k=\lambda^*=F(x_r^*,y^*),\quad T_k=T^*=\Phi(y^*),\quad y^*=x_3^*.
\end{gathered}
\]
In particular, at a 1-cycle, the releasable pool is depleted at each secretion event and replenished between consecutive pulses in such a way that its pre-pulse level is reproduced after one period.

\subsection{The return map and fixed-point equations}

Introduce the matrices
\begin{equation}\label{eq:matrices}
A=
\begin{bmatrix}
-a_1&0&0\\
g_1&-a_2&0\\
0&g_2&-a_3
\end{bmatrix},
B=
\begin{bmatrix}
1\\0\\0
\end{bmatrix},
C^\top=
\begin{bmatrix}
0\\0\\1
\end{bmatrix},
\end{equation}
so that the continuous-time dynamics~\eqref{eq:model_c} can be written as
\[
\dot X=AX,\qquad y=CX.
\]

Given a pre-pulse state $\bar X_k=(x_{r,k},X_k^{\top})^{\top}$, the state variables immediately after the pulse are
\[
x_r(t_k^+)=x_{r,k}-\lambda_k,
\qquad
X(t_k^+)=X_k+\beta B\lambda_k.
\]

Their evolution over the subsequent interval $[t_k,t_{k+1})$ gives the
pre-pulse state at the next release event. 
According to~\eqref{eq:lambda-T-k}, the sequence of pre-pulse states satisfies
\begin{equation}\label{eq:return-dynamics}
    \bar X_{k+1}=\mathcal Q(\bar X_k),
\end{equation}
where the pulse-to-pulse, or return, map
\[
\mathcal Q:\mathbb R_+^4\to\mathbb R_+^4,
\]
is defined by 
\begin{align}
&\mathcal Q(\bar X)
=
\begin{bmatrix}
\e^{-a_r\Phi(y)}
\bigl(x_r-F(x_r,y)\bigr)
+\dfrac{v}{a_r}
\bigl(1-\e^{-a_r\Phi(y)}\bigr)
\\[2mm]
\e^{A\Phi(y)}
\bigl(X+\beta B F(x_r,y)\bigr)
\end{bmatrix}.\label{eq:return-map}
\end{align}
Under admissibility condition~\eqref{eq:F-admissible}, the jump
preserves nonnegativity of the releasable pool $x_r-F(x_r,y)\geq 0$, and therefore
$\mathcal Q$ leaves the nonnegative orthant invariant.

Hence, a 1-cycle corresponds to a fixed point
\[
\bar X^*=\mathcal Q(\bar X^*),
\]
that can be rewritten as a system of two equations
\begin{align}
x_r^*
&=
\e^{-a_rT^*}(x_r^*-\lambda^*)
+\frac{v}{a_r}(1-\e^{-a_rT^*}),
\label{eq:fp-pool}
\\
X^*
&=
\e^{AT^*}(X^*+\beta B\lambda^*),\;\;y^*=CX^*,
\label{eq:fp-goodwin}\\
\lambda^*&=F(x_r^*,y^*),\label{eq:fp-lam}\\
T^*&=\Phi(y^*).\label{eq:fp-T}
\end{align}

\subsubsection*{Scalar fixed-point equation}
One may notice that, similar to the fixed-point equations for the 1-cycle in IGO~\cite{Aut09,PRM24}, the latter equations can further be reduced to scalar equations with respect to the single scalar variable $y^*$. 
Notice first that, since $A$ is Hurwitz, the equations~\eqref{eq:fp-goodwin} entail that
\begin{equation}\label{eq:Xy-fixed}
\begin{gathered}
X^*
=
\beta\lambda^*
(\e^{-AT^*}-I)^{-1}B,\\
y^*
=
\beta\lambda^* R(T^*),
\\
R(\theta)\triangleq
C(\e^{-A\theta}-I)^{-1}B.
\end{gathered}
\end{equation}
In view of~\eqref{eq:fp-T}, $\lambda^*$ and $y^*$ are related by the equation
\begin{equation}\label{eq:Lambda}
\lambda^*=\Lambda(y^*),\quad \Lambda(\theta)
\triangleq
\frac{\theta}{\beta R(\Phi(\theta))}.
\end{equation}
Solving~\eqref{eq:fp-pool} with respect to $x_r^*$ and substituting~\eqref{eq:fp-T} and~\eqref{eq:Lambda}, one finally expresses $x_r^*$ as a function of $y^*$:
\begin{equation}\label{eq:xr-fixed}
\begin{gathered}
x_r^*
=
\frac{v}{a_r}
-
\frac{\lambda^*}{\e^{a_rT^*}-1}=P(y^*),\\
P(\theta)
\triangleq
\frac{v}{a_r}
-
\frac{\Lambda(\theta)}
     {\e^{a_r\Phi(\theta)}-1}.
\end{gathered}
\end{equation}
Notice that the function $P(\cdot)$ can attain negative values, so formally $x_r^*$ found in this way can be infeasible. If $P(y^*)\geq 0$, then
$F(P(y^*),y^*)$ is well-defined and
\begin{equation}\label{eq:scalar-fp}
F(P(y^*),y^*)=\Lambda(y^*).
\end{equation}

The above reduction is reversible, i.e., positive\footnote{Notice that $y^*=0$ cannot be a solution to~\eqref{eq:scalar-fp}, since $F(P(0),0)=F(v/a_r,0)>0=\Lambda(0)$ in view of~\eqref{eq:F-admissible}. Hence, we only consider positive solutions to~\eqref{eq:scalar-fp}, producing strictly positive 1-cycles.} 1-cycles can be characterized completely in terms of the equation \eqref{eq:scalar-fp}. The following proposition makes this relation precise.
\begin{proposition}\label{prop:scalar-eq}
Assume that $\Phi(\theta)>0$ for all $\theta>0$ and~\eqref{eq:F-admissible} holds.
Let $y^*>0$ satisfy the conditions\footnote{As discussed before, $P(y^*)\geq 0$ is the necessary feasibility condition. Furthermore, $P(y^*)=0$ is impossible when $y^*>0$, since $F(0,y^*)=0<\Lambda(y^*)$ in view of~\eqref{eq:F-admissible} and~\eqref{eq:Lambda}.} $P(y^*)>0$ and~\eqref{eq:scalar-fp}.
Then,  system~\eqref{eq:pool},~\eqref{eq:model_c} has a strictly positive 1-cycle, corresponding to the fixed point of~\eqref{eq:return-dynamics} with
\begin{equation}\label{eq:cycle-from-theta}
\begin{gathered}
T^*=\Phi(y^*),\,
\lambda^*=\Lambda(y^*),
\,
x_r^*=P(y^*),\\
X^*=\beta\lambda^*(\e^{-AT^*}-I)^{-1}B.
\end{gathered}
\end{equation}
Conversely, every positive 1-cycle gives rise to a solution $y^*>0$ of~\eqref{eq:scalar-fp} satisfying $P(y^*)>0$.
\end{proposition}

\subsection{Existence and uniqueness of 1-cycle}

The following theorem is the main result of this paper.
\begin{theorem}\label{thm:exist-unique}
Assume that $\Phi:\mathbb R_+\to(0,\infty)$ is continuous and
nondecreasing, and that $F:\mathbb R_+^2\to\mathbb R_+$ is continuous
and satisfies~\eqref{eq:F-admissible}. 
Then system  \eqref{eq:model_c}, \eqref{eq:timing}, \eqref{eq:lambda} admits at least one positive 1-cycle. 
If, in addition, $F$ is continuously differentiable and satisfies~\eqref{eq:F-monotone}, then the positive 1-cycle is unique; the cycle characteristics $y^*,\lambda^*,T^*$, considered as functions of the pool parameters $a_r$ and $v$, satisfy
the monotonicity properties:
\begin{enumerate}
\item $y^*$, $\lambda^*$, and $T^*$ are nondecreasing functions of $v$ for each fixed $a_r>0$;
\item $y^*$, $\lambda^*$, and $T^*$ are nonincreasing functions of $a_r$ for each fixed $v>0$.
\end{enumerate}
If $F$ is strictly increasing with respect to its first argument (that is, $F'_{\xi}(\xi,\theta)>0$),
then the monotonicity of $y^*$ and $\lambda^*$ in both statements is
strict. If, in addition, $\Phi$ is strictly increasing, then the
monotonicity of $T^*$ is strict as well.
\end{theorem}

\begin{remark}[Effect of reduced hormone production]
The monotonicity result provides a simple interpretation of changes in
the hormone production rate $v$. If $v$ decreases while the degradation
rate $a_r$ and the feedback functions $F,\Phi$ remain unchanged, then
the steady-state output $y^*$, the pulse amplitude $\lambda^*$, and the
inter-pulse interval $T^*$ all decrease. Thus, a reduction in hormone
production leads to weaker but more frequent release events. 

The shift toward a lower amplitude and higher frequency of GnRH and downstream LH pulses is a classic feature of male reproductive aging. Research shows a 30\% to 50\% increase in GnRH pulse frequency in older men compared to young men \cite{VKL09}.  The mass and size of each LH burst decrease significantly, leading to lower amplitude  pulses.
\end{remark}

\subsection{Orbital stability}

Assuming that both $F$ and $\Phi$ are continuously differentiable,
to test the (local) orbital stability of the 1-cycle, it suffices to check whether the Jacobian matrix $\mathcal Q'(\bar X^*)$, computed at the corresponding fixed point, is Schur stable~\cite{Churilov2020}. The following proposition produces an explicit form of the Jacobian.
\begin{proposition}\label{prop:Jacobian}
Suppose that $F$ and $\Phi$ are continuously differentiable. Introduce
\begin{equation}\label{eq:notation-jacob}
\begin{gathered}
F_\xi^*\triangleq F_\xi(x_r^*,y^*),\;
F_\theta^*\triangleq F_\theta(x_r^*,y^*),\;
\Phi^*_y \triangleq\Phi'(y^*),\\
E_r\triangleq\e^{-a_rT^*},\quad E\triangleq\e^{AT^*}.
\end{gathered}
\end{equation}
The Jacobian of the return map at the fixed point $\bar X^*$ is
\begin{equation}\label{eq:Jacobian}
\begin{aligned}
\mathcal Q'&(\bar X^*)
={}
\begin{bmatrix}
E_r&0\\
0&E
\end{bmatrix}+\\
&+
\begin{bmatrix}
-E_r\\
\beta EB
\end{bmatrix}
\begin{bmatrix}
F_\xi^*&F_\theta^*C
\end{bmatrix}
+
\begin{bmatrix}
v-a_rx_r^*\\
AX^*
\end{bmatrix}
\begin{bmatrix}
0&\Phi_y^*C
\end{bmatrix}.
\end{aligned}
\end{equation}
\end{proposition}

\section{Simulation}\label{sec:sim}
To illustrate the theoretical results of the paper, consider a modification of the Te regulation model in \cite{Aut09}.
In \eqref{eq:matrices}, the constants $a_i, i=1,2,3$ correspond to the degradation rates of GnRH, LH, and Te, respectively. According to the biological data in Veldhuis and Keenan (1998),
\begin{align*}
0.23~\mathrm{min}^{-1} &\le  a_1 \le 0.69~\mathrm{min}^{-1},\\
0.0087~\mathrm{min}^{-1} &\le  a_2 \le  0.014~\mathrm{min}^{-1}, \\
0.00693~\mathrm{min}^{-1} &\le  a_3 \le  0.0693~\mathrm{min}^{-1}.
\end{align*}
Assume $g_1 = 0.1$, $g_2 = 0.15$. Since continuous dynamics \eqref{eq:model_c} are linear, these gains cannot be distinguished from each other. 
The rate constants are chosen as $a_1=0.46$ (in the portal blood compartment), $a_2=0.0114$, and $a_3=0.0115$. 
When stored inside neuronal vesicles within the hypothalamus, GnRH is chemically protected from enzymatic degradation and can persist  for days to weeks until an action potential triggers its exocytosis. 

The modulation functions in the original formulation of the IGO \cite{MCS06,Aut09} were parametrized as
\[
\Phi(y)= k_1+k_2\frac{(y/h)^2}{1+(y/h)^2},
\]
\[
F(y)= k_3+\frac{k_4}{1+(y/h)^2}.
\]
for positive parameters $k_1,k_2,k_3,k_4,h$. Then, for $y\in \lbrack 0,\infty)$
\[
k_1\le\Phi(y)< k_1+k_2, \quad k_3 < F(y) \le k_3+k_4.
\]
The frequency modulation function is kept the same as in \cite{Aut09}, with $k_1=60$, $k_2=40$.
For the model with releasable pool introduced in the present paper, it applies
\[
0 \le F(y) \le x_r.
\]
Then the amplitude modulation function is selected as
\[
F(x_r,y)=\frac{k_3x_r}{1+(y/h)^2}
\]
and the free coefficient is $0<k_3\le 1$. It thus applies that $\forall y: F(0,y)=0$ and $F(x_r,0)=k_3x_r$.

\begin{figure}
\centering
\includegraphics[width=0.8\linewidth]{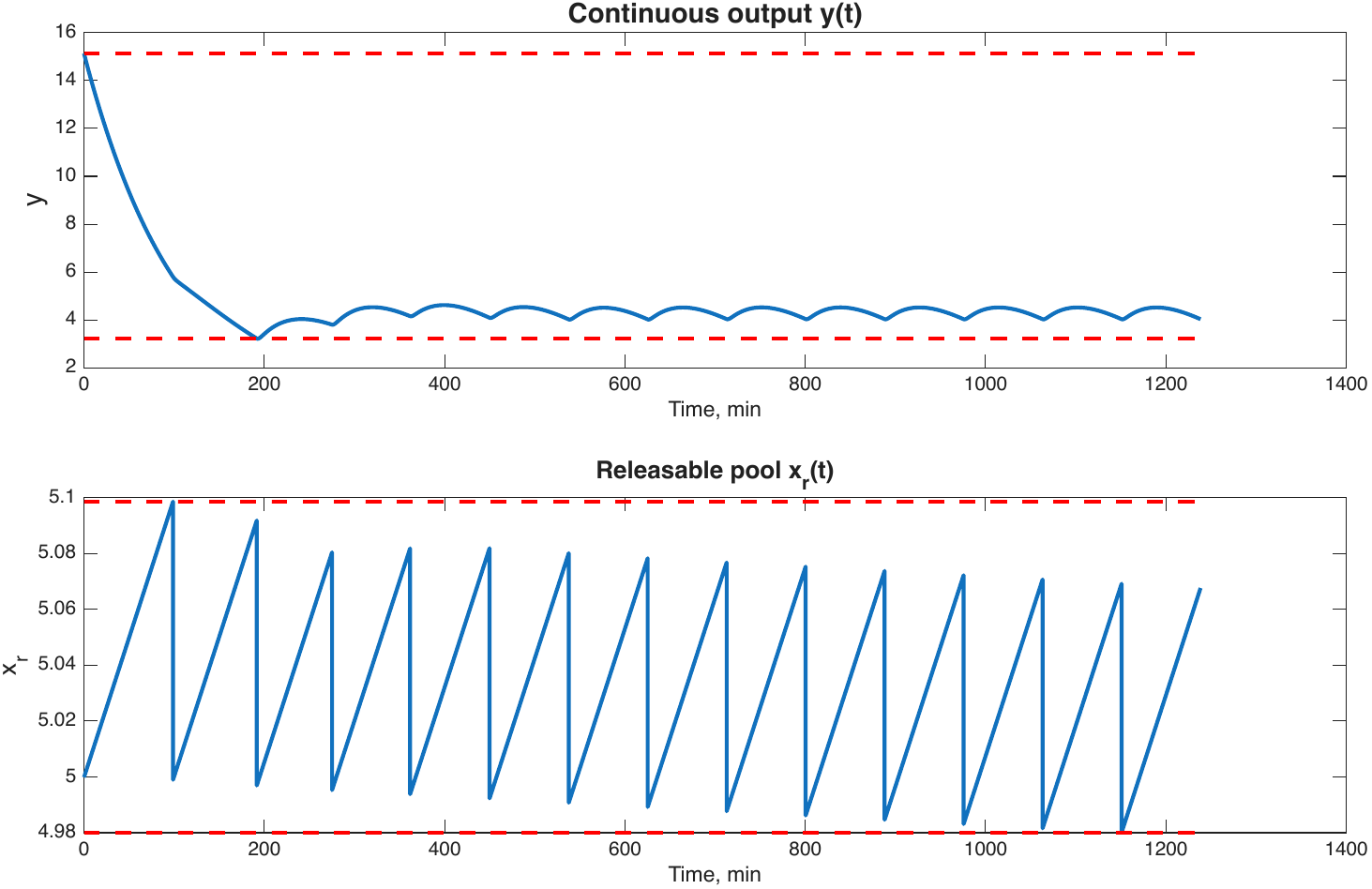}
\caption{Transient response of the Te  model with releasable hormone pool. Red dashed lines show the greatest and least value of each signal and provided for reference. Upper plot: the system output (Te concentration) converging to a 1-cycle. Lower plot: the releasable hormone pool level. }\label{fig:transient}
\end{figure}

\begin{figure}
\centering
\includegraphics[width=0.8\linewidth]{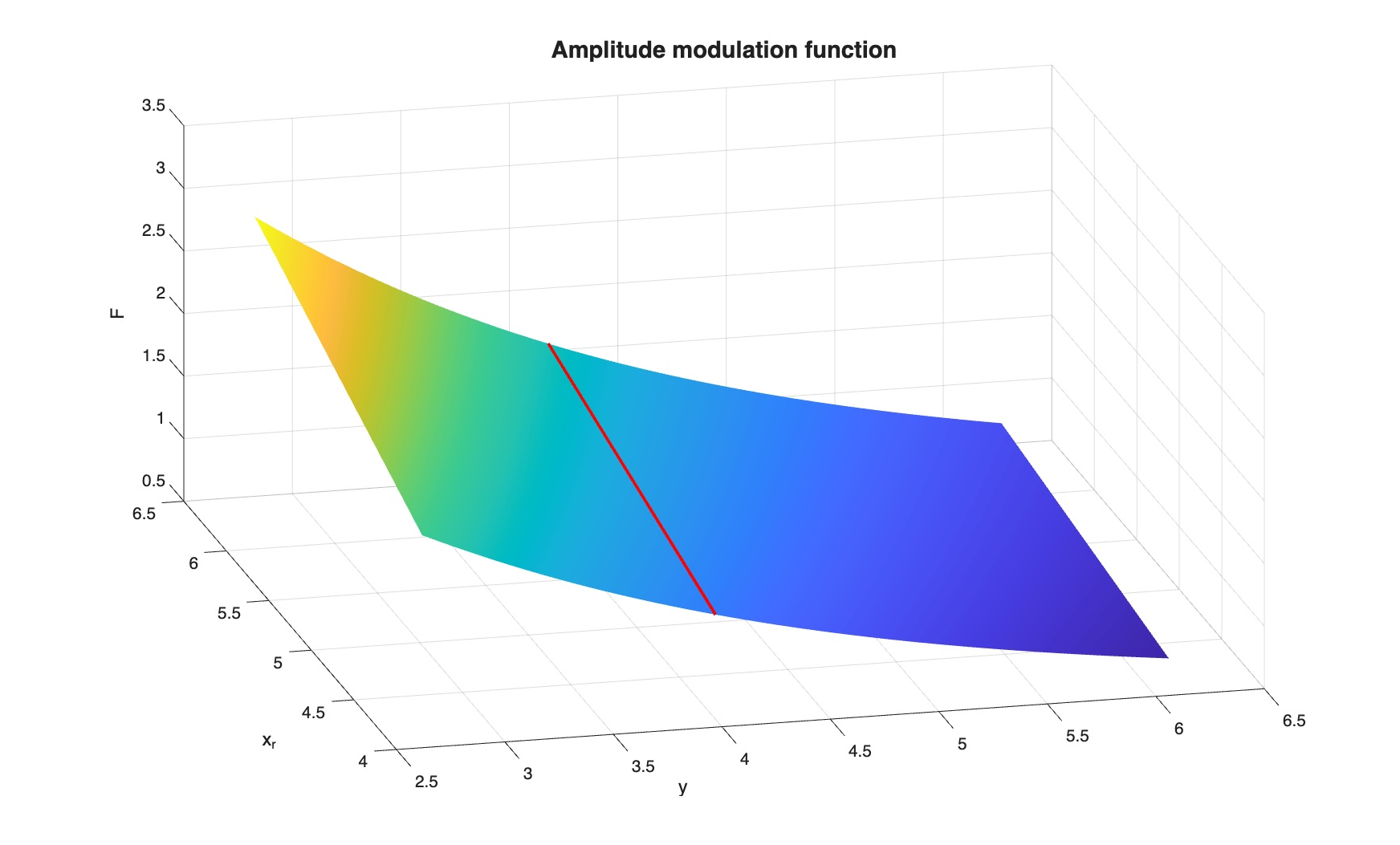}
\caption{Amplitude modulation function. Red line: restriction of $F$ to the fixed point value of the output $y$.}\label{fig:F}
\end{figure}

The shape of the modulation function for the relevant values of the arguments is depicted in Fig.~\ref{fig:F}. As seen from the expression for $F$, it depends linearly on $x_r$, increases in $x_r$ and decreases with $y$. 
This property is also visible in a transient response of the closed-loop system described by \eqref{eq:model_c},\eqref{eq:timing},\eqref{eq:lambda}, Fig.~\ref{fig:transient}. Indeed, the envelope of the $x_r(t)$ signal also decreases linearly over time, signaling insufficient production of GnRH. Although being less apparent, the time intervals between the feedback firings $T_k$ become also shorter (declining from $98.7661$ to $87.6058$), thus implying a rising impulse frequency. This example illustrates the result of Theorem~\ref{thm:exist-unique} on the feedback firing pattern under decreasing level of the releasable hormone pool.

\section*{CONCLUSIONS}
 A well-established and experimentally validated model describing pulsatile endocrine secretion by means of pulse-modulated feedback is extended with a variable representing a releasable hormone pool. It is proven that the essential properties of 1-cycles in the Impulsive Goodwin's Oscillator are preserved in the extended model, despite a higher differential order of the continuous part and the introduction of an amplitude modulation function depending on multiple arguments. The model is shown to reproduce a specific dynamical phenomenon clinically observed in testosterone regulation in aging men consisting of a pulsatile secretion pattern of lower amplitude and higher frequency.



\section{Proofs}\label{sec:proofs}

\subsection{Preliminaries}
\paragraph{Divided differences and the Opitz formula} The first divided difference (DD) of a function $h(\cdot)$ is defined as
\[
h\lbrack x_0, x_1 \rbrack \triangleq \frac{h(x_1)-h(x_0)}{x_1-x_0},\quad x_0\ne x_1.
\]
and higher-order DDs are defined recursively for mutually different arguments $x_0,\ldots,x_k$ by 
\[
h\lbrack x_0, \dots, x_k \rbrack = \frac{h\lbrack x_1, \dots, x_k\rbrack-h\lbrack x_0, \dots, x_{k-1}\rbrack }{x_k-x_0}.
\]
The mean-value theorem for DDs~\cite{DeBoor2005} states that if
$h\in C^k([m,M]\to\mathbb{R})$, where $m=\min_i x_i$ and $M=\max_i x_i$, then a point $\zeta\in [m,M]$ exists such
that
\begin{equation}\label{eq:mean-val}
h[x_0, \dots, x_k] = \frac{1}{k!}h^{(k)}(\zeta).
\end{equation}

In the proofs, we will use the special case of the Opitz formula~\cite{DeBoor2005,PRM24}. Given a function $f$, which is complex-analytic in a vicinity of the spectrum $\{-a_1,-a_2,-a_3\}$ of the matrix $A$ in~\eqref{eq:matrices}, its value on the matrix is found as
\begin{equation*}
f(A)=
\begin{bmatrix}
f(-a_1) & 0 & 0\\
g_1f[-a_1,-a_2] & f(-a_2) & 0\\
g_1g_2f[-a_1,-a_2,-a_3] & g_2f[-a_2,-a_3] & f(-a_3)
\end{bmatrix}.
\end{equation*}
Applying the formula to $TA$, $T>0$, amounts to replacing $a_i$ and
$g_i$ by $a_iT$ and $g_iT$, respectively. In particular, the
bottom-left entry of $f(TA)$ is
\begin{equation}\label{eq:opitz31}
Cf(TA)B=g_1g_2T^2f[-a_1T,-a_2T,-a_3T].
\end{equation}
The mean-value theorem~\eqref{eq:mean-val} yields the
following corollary.
\begin{corollary}\label{cor:opitz-convex}
Let $f\in C^2((-\infty,0))$. If $f''(x)\geq0$ for all
$x<0$, then $Cf(TA)B\geq0$ for all $T>0$.
If $f''(x)>0$ for all $x<0$, then $Cf(TA)B>0$.
\end{corollary}

\paragraph{Some properties of the function $R(T)$}

\begin{figure}
\centering
\includegraphics[width=0.8\linewidth]{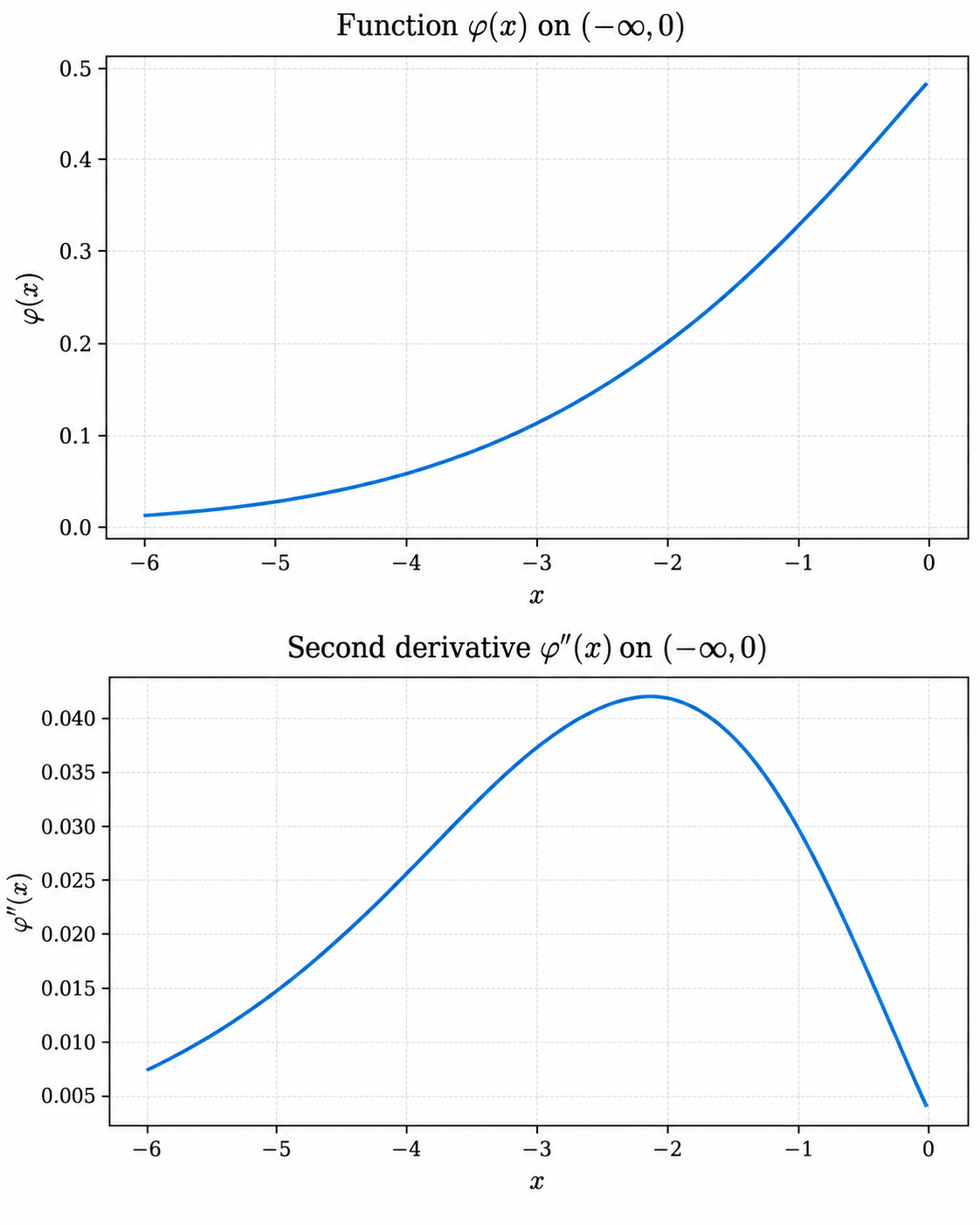}
\caption{Convexity of $\varphi$ from~\eqref{eq:varphi}}\label{fig:convex}
\end{figure}

As was shown in~\cite[Appendix~A]{Aut09}, the function $R(T)$ is positive and decreasing on $(0,+\infty)$. The following lemma establishes additional properties to be used in the further analysis.
\begin{lemma}\label{lem:r}
The function $R(T)=C(\e^{-TA}-I)^{-1}B$ exhibits the following properties:
\begin{enumerate}
\item For every $T>0$, the inequalities hold
\begin{equation}\label{eq:R-strong-monotonicity}
   -T R'(T)>R(T)>0.
\end{equation}
In particular, $R'(T)<0$ and $(TR(T))'=R(T)+TR'(T)<0$, i.e., the functions $R(T)$ and $TR(T)$ are strictly decreasing on $(0,\infty)$.

\item For every $a>0$, the function
\begin{equation}\label{eq:Da}
   D_a(T)\triangleq (1-\e^{-aT})R(T),
\end{equation}
is strictly decreasing on $(0,\infty)$.

\end{enumerate}
\end{lemma}
\vskip1mm
\begin{proof}
The right inequality in~\eqref{eq:R-strong-monotonicity} $R(T)>0$ has been already proved in~\cite{Aut09}. To prove the left inequality, notice that
$R'(T)=CA\e^{-TA}(\e^{-TA}-I)^{-2}B$, and hence
\[
-TR'(T)-R(T)=C\varphi(TA)B,
\]
where the function $\varphi(x)$ is defined as
\begin{equation}\label{eq:varphi}
\varphi(x)\triangleq-\frac{x\e^{-x}}{(\e^{-x}-1)^2}-\frac{1}{\e^{-x}-1}=\frac{1-(x+1)\e^{-x}}{(\e^{-x}-1)^2}.
\end{equation}
It can be checked that $\varphi$ is strictly convex on $(-\infty,0)$ since $\varphi''(x)>0$ for $x<0$ as illustrated in Fig.~\ref{fig:convex}.
By Corollary~\ref{cor:opitz-convex}, this implies that $-TR'(T)-R(T)>0$ for $T>0$, which concludes the proof of statement 1).

To prove statement 2), notice that, for each $a>0$ and $T>0$, one has $\e^{aT}>1+aT$, and hence
\[
-\frac{R'(T)}{R(T)}\geq\frac{1}{T}>\frac{a}{\e^{aT}-1}=\frac{a\e^{-aT}}{1-\e^{-aT}},
\]
entailing that
\[
\begin{aligned}
D_a'(T)&=a\e^{-aT}R(T)+(1-\e^{-aT})R'(T)\\&=R(T)(1-\e^{-aT})\left(\frac{a\e^{-aT}}{1-\e^{-aT}}+\frac{R'(T)}{R(T)}\right)<0.
\end{aligned}
\]
This proves that $D_a$ is strictly decreasing on $(0,\infty)$.
\end{proof}

\subsection{Proof of Proposition~\ref{prop:scalar-eq}}

Let first $y^*>0$ satisfy~\eqref{eq:scalar-fp} and
$P(y^*)>0$, and define $T^*$, $\lambda^*$, $x_r^*$, and
$X^*$ by~\eqref{eq:cycle-from-theta}. Since $\Phi(y^*)>0$ and $R(T)>0$ for $T>0$ (Statement~1) in Lemma~\ref{lem:r}), one has
\[
    T^*>0,
    \qquad
    \lambda^*
    =
    \beta^{-1}y^*/R(\Phi(y^*))
    >0.
\]
Applying the Opitz formula to the function $f(x)=(\e^{-x}-1)^{-1}$, which is positive, increasing, and strictly convex for $x<0$, one proves that $X^*>0$. Moreover, $
    CX^*
    =
    \beta\lambda^*
    C(\e^{-AT^*}-I)^{-1}B
    =
    \beta\lambda^*R(T^*)
    =y^*.
$

It remains to prove that equations~\eqref{eq:fp-pool}-\eqref{eq:fp-T} are satisfied. The last equation~\eqref{eq:fp-T} is straightforward.
By~\eqref{eq:scalar-fp},
\[
    F(x_r^*,y^*)
    =
    F(P(y^*),y^*)
    =
    \Lambda(y^*)
    =
    \lambda^*,
\]
so  amplitude-modulation equation~\eqref{eq:fp-lam} is also satisfied. Next, the definition of $X^*$ gives
\[
    (\e^{-AT^*}-I)X^*=\beta\lambda^*B,
\]
which is equivalent to fixed-point equation~\eqref{eq:fp-goodwin}. 
Similarly, from the definition of $P$ in~\eqref{eq:xr-fixed}, one has
\[
    x_r^*
    =
    \frac{v}{a_r}
    -
    \frac{\lambda^*}{\e^{a_rT^*}-1},
\]
which is equivalent to~\eqref{eq:fp-pool}. We have thus constructed a positive fixed point $\bar X^*$
of the return map $Q$. 

Conversely, let $\bar X^*$ be a positive fixed point corresponding to the cascade's output $y^*>0$.
Equations~\eqref{eq:Xy-fixed} and~\eqref{eq:fp-T} imply
$\lambda^*=\Lambda(y^*)$, while~\eqref{eq:fp-pool} gives
$x_r^*=P(y^*)$. Since $x_r^*>0$, one has $P(y^*)>0$, and
\eqref{eq:fp-lam} yields~\eqref{eq:scalar-fp}.\hfill$\blacksquare$

\subsection{Proof of Theorem~\ref{thm:exist-unique}}

Consider the equation
\begin{equation}\label{eq:H}
H(\theta)
\triangleq
F(P(\theta),\theta)-\Lambda(\theta)=0.
\end{equation}
In view of Proposition~\ref{prop:scalar-eq}, positive 1-cycles are in one-to-one correspondence with positive solutions of~\eqref{eq:H},
satisfying the condition $P(\theta)>0$.

\paragraph*{Existence} Recall that, according to Lemma~\ref{lem:r}, $R$ is a decreasing function on $(0,\infty)$, and hence its composition $R\circ\Phi$ with the non-decreasing and positive function $\Phi$ is non-increasing on $[0,\infty)$. Using the definition of $\Lambda(\theta)$ in~\eqref{eq:Lambda}, one proves that this function is strictly increasing on $[0,\infty)$. 

Replacing $R(\Phi(\theta))$ by $D_{a_r}(\Phi(\theta))$, where $D_{a_r}$ is defined in~\eqref{eq:Da}, the same arguments apply
to the function
\begin{equation}\label{eq:S-def}
\begin{aligned}
    S(\theta)
    &\triangleq\frac{\theta}
         {\beta D_{a_r}(\Phi(\theta))}=
    \frac{\Lambda(\theta)}
         {1-\e^{-a_r\Phi(\theta)}}\\
    &=\frac{\Lambda(\theta)\e^{a_r\Phi(\theta)}}
         {\e^{a_r\Phi(\theta)}-1}=\Lambda(\theta)+\frac{\Lambda(\theta)}
         {\e^{a_r\Phi(\theta)}-1},
\end{aligned}
\end{equation}
that is strictly increasing on $[0,\infty)$, because $D_{a_r}$ is decreasing on $(0,\infty)$ (Lemma~\ref{lem:r}). Obviously, $S(0)=0$. Moreover, since $D_{a_r}\circ\Phi$ is non-increasing, one has
\[
S(\theta)
=
\frac{\theta}{\beta D_{a_r}(\Phi(\theta))}
\geq
\frac{\theta}{\beta D_{a_r}(\Phi(0))}
\xrightarrow[\theta\to\infty]{}\infty.
\]
Consequently, there exists a unique $\bar\theta>0$ such that
\begin{equation}\label{eq:theta-bar}
    S(\bar\theta)=\frac{v}{a_r}.
\end{equation}
Notice that the function $P$ in~\eqref{eq:xr-fixed} can be written as
\begin{equation}\label{eq:P-Lambda-S}
    P(\theta)
    =
    \frac{v}{a_r}
    -S(\theta)+\Lambda(\theta).
\end{equation}
For any $\theta\in [0,\bar\theta]$, one has $S(\theta)\leq S(\bar\theta)$ and hence
\[
P(\theta)\geq \frac{v}{a_r}
    -S(\bar\theta)+\Lambda(\theta)=\Lambda(\theta)>0.
\]

Using the continuity, we now prove that equation~\eqref{eq:H} has at least one solution on $(0,\bar\theta]$.
Indeed, at $\theta=0$, one has
\[
P(0)=\frac{v}{a_r},
\quad
H(0)
=
F\left(\frac{v}{a_r},0\right)>0,
\]
in view of~\eqref{eq:F-admissible}. At the other endpoint,~\eqref{eq:theta-bar} and
\eqref{eq:P-Lambda-S} give
$P(\bar\theta)=\Lambda(\bar\theta)$. By using condition~\eqref{eq:F-admissible}, one shows that
\[
H(\bar\theta)
=
F(\Lambda(\bar\theta),\bar\theta)
-\Lambda(\bar\theta)
\leq0.
\]
By continuity, $H$ has at least one zero $\theta^*\in(0,\bar\theta]$, satisfying also $P(\theta^*)>0$, proving the existence part of the theorem.

\paragraph*{Uniqueness} We are now going to prove uniqueness of the 1-cycle under additional constraint~\eqref{eq:F-monotone}.
To this end, we show that $H$ is strictly decreasing on its admissible domain -- the set of $\theta\geq 0$ such that $P(\theta)\geq 0$. For $\theta_2>\theta_1$, let 
\[
\Lambda_i\triangleq\Lambda(\theta_i),\,
S_i\triangleq S(\theta_i),\,
P_i\triangleq P(\theta_i),\,
H_i\triangleq H(\theta_i).
\]
The conditions in~\eqref{eq:F-monotone} and the mean value theorem imply that
\[
\begin{aligned}
F(P_2,\theta_2)&-F(P_1,\theta_1)\leq F(P_2,\theta_1)-F(P_1,\theta_1)\\
&=F'_{\xi}(\xi_0,\theta_1)(P_2-P_1)\leq\max(P_2-P_1,0),
\end{aligned}
\]
where $\xi_0$ is some point lying between $P_1$ and $P_2$. By using the definition of $H$ and~\eqref{eq:P-Lambda-S},
one arrives at the inequalities
\[
\begin{aligned}
H_2-H_1&=F(P_2,\theta_2)-F(P_1,\theta_1)-(\Lambda_2-\Lambda_1)\\
&\leq \max(P_2-P_1-(\Lambda_2-\Lambda_1),-(\Lambda_2-\Lambda_1))\\
&=\max(S_1-S_2,\Lambda_1-\Lambda_2)<0,
\end{aligned}
\]
where the last maximum is negative, since both $S(\theta)$ and $\Lambda(\theta)$ are strictly increasing functions and $\theta_2>\theta_1$. Hence,  $H$ is strictly decreasing, and the solution of~\eqref{eq:H} is unique.

\paragraph*{Monotonicity}

To make the dependence on the pool parameters explicit, we write
\[
P(\theta;v,a_r)
\triangleq
\frac{v}{a_r}
-
\frac{\Lambda(\theta)}
     {e^{a_r\Phi(\theta)}-1}.
\]
Thus, $P(\theta;v,a_r)$ is the pre-pulse level of the releasable
pool that would be compatible with a 1-cycle having output
$y^*=\theta$, for the given values of $v$ and $a_r$. The same notation rule is applied to $H$, denoting
\[
H(\theta;v,a_r)
\triangleq
F(P(\theta;v,a_r),\theta)-\Lambda(\theta).
\]

\textbf{Dependence on $v$.} To study the dependence on $v$, fix $a_r$ and let $v_2>v_1$.
For every $\theta\geq0$, the definition of $P$ in~\eqref{eq:xr-fixed} implies
$
P(\theta;v_2,a_r)-P(\theta;v_1,a_r)
=
(v_2-v_1)/a_r>0.
$
Since $F$ is non-decreasing with respect to its first argument,
\[
H(\theta;v_2,a_r)
\geq
H(\theta;v_1,a_r)
\qquad\forall\,\theta\geq0.
\]

Let $\theta_i^*$ be the unique zero of $H(\cdot;v_i,a_r)$.
Then
\[
H(\theta_1^*;v_2,a_r)\geq
H(\theta_1^*;v_1,a_r)=0.
\]
Since $H(\cdot;v_2,a_r)$ is strictly decreasing, its unique zero
$\theta_2^*$ satisfies 
$\theta_2^*\geq\theta_1^*$.
Thus $y^*=\theta^*$ is non-decreasing in $v$.
Since $\Lambda$ is strictly increasing and $\Phi$ is non-decreasing,
\[
\lambda^*=\Lambda(y^*),
\qquad
T^*=\Phi(y^*),
\]
are also non-decreasing in $v$. If $F$ is strictly increasing with respect to its first argument,
then
$
H(\theta;v_2,a_r)>
H(\theta;v_1,a_r),
$
so $\theta_2^*>\theta_1^*$. Hence $y^*$ and $\lambda^*$ are strictly
increasing in $v$. If, in addition, $\Phi$ is strictly increasing,
then $T^*$ is strictly increasing in $v$ as well.

\textbf{Dependence on $a_r$.} Notice first that the function
\[
a_r\mapsto\frac{a_r}{1-\e^{-a_rT}}
\]
is strictly increasing on $(0,\infty)$ for every $T>0$. Indeed,
\[
\frac{\rm d}{\rm d a_r}\frac{a_r}{1-\e^{-a_rT}}
=
\frac{1-(1+a_rT)\e^{-a_rT}}
     {(1-\e^{-a_rT})^2}>0,
\]
since $\e^{a_rT}>1+a_rT$.
Consequently, for every fixed $\theta>0$, 
\[
\widehat S(\theta;a_r)
\triangleq
a_r S_{a_r}(\theta)
=
\frac{a_r\Lambda(\theta)}
     {1-\e^{-a_r\Phi(\theta)}},
\]
is strictly increasing in $a_r$.

Let $a_{r,2}>a_{r,1}$, and let $\theta_i^*$ be the unique
positive solution of~\eqref{eq:H} corresponding to $a_{r,i}$.
By the admissibility condition~\eqref{eq:F-admissible},
$\Lambda(\theta_2^*)
=
F(P(\theta_2^*;v,a_{r,2}),\theta_2^*)
\leq P(\theta_2^*;v,a_{r,2}).
$
This implies the inequality $v-\widehat S(\theta_2^*;a_{r,2})\geq0$, because
\[
P(\theta;v,a_r)
=
\Lambda(\theta)
+
\frac{v-\widehat S(\theta;a_r)}{a_r}.
\]
Since $\widehat S(\theta_2^*;a_r)$ is strictly increasing
in $a_r$, one has
\[
\begin{aligned}
\frac{v-\widehat S(\theta_2^*;a_{r,1})}{a_{r,1}}
&>
\frac{v-\widehat S(\theta_2^*;a_{r,2})}{a_{r,1}}\\
&\geq
\frac{v-\widehat S(\theta_2^*;a_{r,2})}{a_{r,2}}.
\end{aligned}
\]
Therefore,
\[
P(\theta_2^*;v,a_{r,1})
>
P(\theta_2^*;v,a_{r,2})
\geq\Lambda(\theta_2^*)>0.
\]
In particular, both values of $P$ are positive, so the
corresponding values of $H$ are well-defined.
Since $F$ is nondecreasing in its first argument,
\[
H(\theta_2^*;v,a_{r,1})
\geq
H(\theta_2^*;v,a_{r,2})=0.
\]
As $H(\cdot;v,a_{r,1})$ is strictly decreasing and
vanishes at $\theta_1^*$, it follows that
$\theta_2^*\leq\theta_1^*$. Thus $y^*=\theta^*$ is non-increasing in $a_r$.
Since $\Lambda$ is strictly increasing and $\Phi$ is
non-decreasing, the same holds for
$\lambda^*=\Lambda(y^*)$ and $T^*=\Phi(y^*)$.

If $F$ is strictly increasing in its first argument,
then $H(\theta_2^*;v,a_{r,1})>0$, implying
$\theta_2^*<\theta_1^*$. Consequently, $y^*$ and
$\lambda^*$ are strictly decreasing in $a_r$.
If, in addition, $\Phi$ is strictly increasing, then
$T^*$ is strictly decreasing in $a_r$ as well.
\hfill$\blacksquare$

\subsection{Proof of Proposition~\ref{prop:Jacobian}}

\noindent Rewriting the formula for the return map $\mathcal Q(\bar X)$ in~\eqref{eq:return-map} as
\begin{align*}
\begin{bmatrix}
\e^{-a_r\Phi(CX)}
\bigl(x_r-F(x_r,CX)\bigr)
+\dfrac{v}{a_r}
\bigl(1-\e^{-a_r\Phi(CX)}\bigr)
\\[2mm]
e^{A\Phi(CX)}
\bigl(X+\beta B F(x_r,CX)\bigr)
\end{bmatrix},
\end{align*}
we denote the upper scalar block by $\mathcal Q_r$ and the lower three-dimensional block by $\mathcal Q_X$. The derivative at the fixed point will by denoted by the symbol $\vert_*$.

Using the notation in~\eqref{eq:notation-jacob}, differentiation over $x_r$ gives
\[
\begin{aligned}
\frac{\partial Q_r}{\partial x_r}\Big|_*
&=
E_r(1-F_\xi^*)=D_1,\\
\frac{\partial Q_r}{\partial X}\Big|_*
&=
\left(
-E_rF_\theta^*
+
E_r\bigl(v-a_r(x_r^*-\lambda^*)\bigr)\Phi^*_y
\right)C\\
&=\left(
-E_rF_\theta^*
+
(v-a_rx_r^*)\Phi^*_y
\right)C=D_2,
\end{aligned}
\]
where the last equality follows from the fixed-point relation~\eqref{eq:fp-pool}, because
$
E_r\bigl(v-a_r(x_r^*-\lambda^*)\bigr)
=
v-a_rx_r^*.
$

Differentiating the vector-valued map $\mathcal Q_X$, one has
\[
\begin{aligned}
\frac{\partial Q_X}{\partial x_r}\Big|_*
&=
\beta EB F_\xi^*=D_3,\\
\frac{\partial Q_X}{\partial X}\Big|_*
&=
E
+
\left[
\beta EB F_\theta^*
+
AE(X^*+\beta B\lambda^*)\Phi^*_y
\right]C=\\
&=E
+
\left[
\beta EB F_\theta^*
+
AX^*\Phi^*_y
\right]C=D_4.
\end{aligned}
\]
where the last simplification is used on the fixed-point equation~\eqref{eq:fp-goodwin}, entailing $E(X^*+\beta B\lambda^*)=X^*$.

Collecting the four blocks, one obtains the relation 
\[
\mathcal Q'(\bar X^*)=\begin{bmatrix}
D_1&D_2\\
D_3&D_4
\end{bmatrix},
\]
which is equivalent to~\eqref{eq:Jacobian} after regrouping the terms.

\bibliographystyle{IEEEtran}
\bibliography{observer,refs} 

\end{document}